\documentclass[a4paper,aps,twoside,twocolumn,superscriptaddress,showpacs]{revtex4}

\usepackage{epsfig,amsmath,amssymb}

\begin{document}

\title{Ballistic deposition patterns beneath a growing KPZ interface}
% {Beneath KPZ growth: From ``crevices'' in ballistic
%   deposition patterns \\
%   to discrete shocks in Burgers turbulence}

\date{\today}

\author{Konstantin Khanin}
\affiliation{Department of Mathematics, University of Toronto, 100
  St~George Street, Toronto, Ontario M5S 3G3, Canada}
\author{Sergei Nechaev}
\affiliation{LPTMS, Universit\'e Paris Sud, 91405 Orsay Cedex, France}
\affiliation{P.N. Lebedev Physical Institute of the Russian Academy of
  Sciences, 53 Leninski ave., 119991, Moscow, Russia}
\affiliation{J.-V. Poncelet Laboratory, Independent University of
 Moscow, 11 B.~Vlasievski per., 119002 Moscow, Russia}
\author{Gleb Oshanin}
\affiliation{LPTMC, Universit\'e Paris 6, 4 Place Jussieu, 75252
  Paris, France}
\affiliation{J.-V. Poncelet Laboratory, Independent University of
 Moscow, 11 B.~Vlasievski per., 119002 Moscow, Russia}
\author{Andrei Sobolevski}
\affiliation{A.A. Kharkevich Institute for Information Transmission
  Problems of~the~Russian Academy of Sciences,
  19 B.~Karetny per., 127994 Moscow, Russia}
\affiliation{J.-V. Poncelet Laboratory, Independent University of
 Moscow, 11 B.~Vlasievski per., 119002 Moscow, Russia}
\author{Oleg Vasilyev}
\affiliation{Max-Planck-Institut f{\"u}r Metallforschung,
  Heisenbergstr.~3, D-70569 Stuttgart, Germany}
\affiliation{Institut f{\"u}r Theoretische und Angewandte Physik,
  Universit{\"a}t Stuttgart, Pfaffenwaldring 57, D-70569 Stuttgart,
  Germany}

\begin{abstract}
  We consider a $(1 + 1)$ dimensional ballistic deposition process
  with next-nearest neighbor interaction, which belongs to the KPZ
  universality class, and introduce for this discrete model a
  variational formulation similar to that for the randomly forced
  continuous Burgers equation.  This allows to identify the
  characteristic structures in the bulk of a growing aggregate
  (``clusters'' and ``crevices'') with minimizers and shocks in the
  Burgers turbulence, and to introduce a new kind of equipped Airy
  process for ballistic growth.  We dub it the ``hairy Airy process''
  and investigate its statistics numerically.  We also identify
  scaling laws that characterize the ballistic deposition patterns in
  the bulk: the law of ``thinning'' of the forest of clusters with
  increasing height, the law of transversal fluctuations of cluster
  boundaries, and the size distribution of clusters.  The
  corresponding critical exponents are determined exactly based on the
  analogy with the Burgers turbulence and simple scaling
  considerations.
\end{abstract}


\maketitle

\section{Introduction}
\label{sect:1}

Over the past few decades, the problem of growth of aggregates by
sequential stochastic deposition developed into one of the most
extensively studied topics in statistical physics \cite{zhang}.  Much
effort has been put into theoretical, numerical, and experimental
investigation of the resulting patterns.  Several theoretical models
have been proposed, including the famous Kardar--Parisi--Zhang (KPZ)
\cite{KPZ} and Edwards--Wilkinson (EW) \cite{EW} models, the
Restricted Solid-on-Solid (RSOS) \cite{rsos} and Eden \cite{eden}
models, the models of Molecular Beam Epitaxy (MBE) \cite{MBE},
Polynuclear Growth (PNG) \cite{M,PS,PNG,BR,J}, and several
ramifications of the Ballistic Deposition (BD) model
\cite{Mand,MRSB,KM,BMW}.  Within the latter, in the simplest setting,
one assumes that elementary units (``particles'') follow ballistic
trajectories in space and adhere sequentially to a growing aggregate
(``heap'').  Despite its extremely transparent geometric formulation,
the problem of stochastic growth still is one of the most puzzling
problems in statistical mechanics.

The available theoretical analysis of stochastic deposition focuses
almost exclusively on the enveloping surface $h(x, t)$, involving a
statistical study of its height distribution and the corresponding
scaling exponents.  Here we quote just a few prominent results.  The
essential scaling relations characterizing the growing aggregate are
\begin{eqnarray*}
  \langle \mathop{\mathrm{Var}}\tilde h(x,t)\rangle^{1/2}
  &\sim& t^{1/3}, \\
  \langle\tilde{h}(x, t)\,\tilde{h}(x + t^{2/3}l, t)\rangle
  - \langle\tilde{h}\rangle^2 &\sim& t^{2/3} F(l).
\end{eqnarray*}
Here $\tilde h(x, t) = h(x, t) - ct$, $c=\lim_{t\to\infty} t^{-1} h(x,
t)$ is an average speed of growth, and $F(l)$ is a rescaled
correlation function.  The exponents $1/3$ and $2/3$ were determined
already in \cite{KPZ} for the KPZ model and then observed in a variety
of other growth models.  Then in Refs~\cite{BDJ,PS} it was realized
that the distribution of a rescaled PNG height $t^{-1/3}(h(0, t) -
2t)$ converges as $t\to\infty$ to the Tracy--Widom distribution
\cite{TW} for the Gaussian unitary ensemble (GUE), which appears in the theory of random matrices.  Moreover,
the full rescaled PNG surface $t^{-1/3}(h(x t^{2/3}, t) - 2t) + x^2$
converges to a version of the Airy stochastic process $Airy_2(x)$
\cite{PNG} whose one-point distributions are precisely Tracy--Widom.
Distribution of maximal heights of the $(1 + 1)$ dimensional
Edwards--Wilkinson and KPZ interfaces has been determined exactly in
Ref.~\cite{SC}.

It should be pointed out that the BD model considered below involves
the \emph{point-to-line} last-passage percolation while PNG
corresponds to the \emph{point-to-point} setting~\cite{J}.
Correspondingly the limit processes are different: it is $Airy_1$ for
the point-to-line and $Airy_2$ for the point-to-point.  Note also that
the one-point distribution for the $Airy_1$ process is given by the
Gaussian orthogonal ensemble (GOE) distribution rather than the GUE distribution, which corresponds
to~$Airy_2$.

Since a similar convergence to Airy processes is observed in other
growth models such as TASEP \cite{sasam}, it is becoming customary to
speak of ``the KPZ universality class'' whenever such limit behavior
is present.  For example, a KPZ scaling has been shown in Refs\
\cite{scaling1,scaling2,scaling3} for the BD model in the
thermodynamic limit.

Much less is known, however, about the structure of BD patterns
\emph{beneath} the enveloping surface.  Here is just one puzzle:
analytic arguments \cite{Vershik.A:2000a} predict that the expected
density $\rho_{\rm surf}$ of local surface maxima in a $(1 + 1)$
dimensional ballistically growing heap is $\rho_{\rm surf}=1/3$,
whereas extensive numerical simulations show that the mean bulk
density, $\rho_{\rm bulk}$, of the $(1 + 1)$ dimensional heap is about
$0.25 \approx 1/4$.  To date there is no satisfactory quantitative
explanation of this mismatch.

\begin{figure}
\epsfig{file=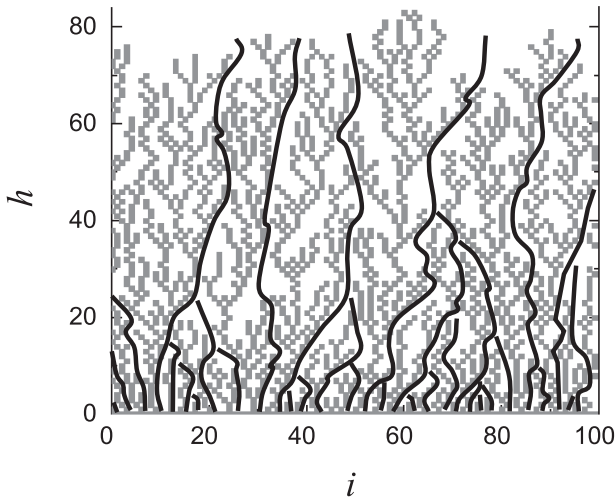, width=8cm}
\caption{Snapshot of a heap obtained by ballistic deposition of
  $N=2000$ particles in a periodic box of size $L=100$ with
  next-nearest-neighbor interactions.  Black lines trace the channels
  (``crevices'') between adjacent clusters.}
\label{fig:1}
\end{figure}

The statistics of the growing heap are determined by its striking
internal structure, revealed in numerical simulations as well as in
the recent experimental analysis of electrochemically formed silver
branched patterns \cite{silver}.  This structure consists of a
``forest'' formed by tree--like clusters of different size, which are
separated by a dual network of tree--like channels or ``crevices''
(Fig.~\ref{fig:1}).

As the heap grows, clusters randomly collect particles and thus spread
and isolate their neighbors from the ``rainfall'' of incident
particles, suffocating their growth.  Consequently the number of
clusters present at height $h$ in a growing aggregate is a decreasing
function of $h$.  We remark that this ``suffocation'' mechanism, as
well as the growth patterns in the BD model, bear certain similarity
to those observed in diffusion limited aggregation in a hard--core
lattice gas on a zero--temperature boundary~\cite{oshanin3}, although
the two models belong to different universality classes and their
quantitative behaviors are in no direct correspondence.

In the present work we undertake investigation of clusters and
crevices based on a novel systematic analogy with turbulent structures
in randomly forced Hamilton--Jacobi equations.  This allows us to
conclude that BD belongs to a large group of models within the KPZ
universality class, such as PNG model, TASEP, and others.  It turns
out that BD like other models mentioned above admits a variational
formulation.  Moreover the analogy with Hamilton--Jacobi dynamics
enables us to suggest a novel concept of \emph{equipped Airy process},
a buildup on top of classical Airy processes which also takes into
account the geometrical structure of the optimal paths (maximizers of
the action, see Section~\ref{sect:3}).  The random field of optimal
paths arises naturally in the context of stochastically forced Burgers
equation \cite{EKMS,khanin_bec}.

In a recent experimental work \cite{silver} the size distribution $P(m)$ of frozen structures
formed by electrochemically grown silver branching patterns has been analyzed.  The authors found
that the probability $P(m)$ to have a cluster of size $m$ exhibits scale invariance, i.e.\ $P(m)
\sim m^{\tau}$, with a critical exponent $\tau=1.37 \pm 0.04$.  In our work we compute this
exponent analytically ($\tau=7/5$) and show that the scaling behavior conjectured in \cite{silver}
actually holds as well as two other power laws governing the ``thinning'' of the forest of clusters
with increasing height and the transversal fluctuations of the cluster boundaries.

The paper is organized as follows.  In Section \ref{sect:2} we specify
the model and define its main structural features.  Section
\ref{sect:3} contains an analysis of the structural similarity of BD
patterns to ``minimizers'' and ``shocks'' in the Burgers turbulence
\cite{khanin_bec}, based on the common variational formulation of the
two models.  In Section \ref{sect:airy} we discuss the KPZ scaling in
the BD model and introduce the notion of an equipped Airy process.
Building on these developments, in Section \ref{sect:scaling} we
compute the main scaling exponents of the BD model.  Section
\ref{sect:conclusion} contains concluding remarks and outlook for
future work.

\section{The model and basic definitions}
\label{sect:2}

\subsection{The NNN ballistic deposition model}
\label{sect:model}

A standard $(1 + 1)$ dimensional BD model with next-nearest-neighbor
(NNN) interactions can be formulated as follows (see also Refs\
\cite{scaling1,scaling2,scaling3}).  Consider a box divided into $L$
columns of unit width each, enumerated with index $i$ ($i = 1, 2,
\dots, L$).  For simplicity we assume the periodic boundary
conditions, so that the leftmost and the rightmost columns are
neighbors, and identify the index value $0$ with~$L$.

At the initial time $t=0$ the system is empty.  Then, at each time step $t = 1, 2, \dots, t_{\rm
max}$, an elementary unit (``particle'') of height~$\ell$ and width $1$ is deposited at a column
$i(t)$ chosen randomly with uniform distribution.
\begin{subequations}
  Define
  \begin{equation}
    \eta_i(t) = \begin{cases}
      1, & i = i(t), \\
      0, & i \neq i(t).
    \end{cases}
    \label{eq:1a}
  \end{equation}
  As shown in Fig.~\ref{fig:1}, particles deposited in adjacent
  columns interact in such a way that they can only touch each other
  at corners or at top and bottom, but never along their vertical
  sides.  Let the height of column $i$ at time $t - 1$ be~$h_i(t -
  1)$.  Upon adding a particle it changes according to
  \begin{equation}
    h_i(t)=\begin{cases}
      \max\{h_{i-1}(t - 1), h_i(t - 1),
      \text{\rlap{$h_{i+1}(t - 1)\}+\ell,$}} \\
      {} & \eta_i(t) = 1, \\[2ex]
      \mathstrut\qquad h_i(t - 1), & \eta_i(t) = 0.
    \end{cases}
    \label{eq:1b}
  \end{equation}
\end{subequations}
This dynamics is supplemented with the initial condition $h_i(0)\equiv
0$ for all $1\le i \le L$.  Eqs~\eqref{eq:1a}, \eqref{eq:1b}
completely describe updating rules for the NNN discrete ballistic
deposition.

We will use Eq.~\eqref{eq:1b} represented in a different form.  Define the ``thin'' and ``thick''
discrete ``$\delta$--functions''
\begin{equation}
  L_{k, i}^0 = \begin{cases} \infty & |k - i| > 0, \\
    0 & |k - i| = 0, \end{cases} \quad
  L_{k, i}^1 = \begin{cases} \infty & |k-i|>1, \\
    0 & |k-i|\leq 1 \end{cases}.
  \label{eq:idr}
\end{equation}
Consider first the trivial dynamics described by the equation $h_i(t) = h_i(t - 1)$.  It can be
rewritten as $h_i(t) = \max\limits_k\, [h_k(t - 1) - L_{k, i}^0]$: indeed, $\max\limits_{k \neq
i}\, [h_k(t - 1) -\infty]\equiv -\infty$ and therefore $h_i(t) = \max\, \{h_i(t - 1), -\infty\} =
h_i(t - 1)$. It is now clear that the stochastic equation~\eqref{eq:1b} can be recast in the form
\begin{equation}
  h_i(t) = \max_{k}\, [h_k(t - 1) - L_{k,i}^{\eta_i(t)}] + \ell\eta_{i}(t).
  \label{eq:nnn}
\end{equation}

This dynamics should be compared with the commonly used discrete
equation with ``additive noise'' describing the $(1 + 1)$ dimensional
polynuclear growth \cite{PNG}, which in our notation takes the form
\begin{equation}
  \tilde h_i(t)
  = \max_{k}\, [\tilde h_k(t - 1) - L_{k,i}^1] + \ell\eta_i(t).
\label{eq:png}1`
\end{equation}
According to Eq.~\eqref{eq:nnn}, the height $h_i$ remains unchanged (quenched) if nothing is
deposited to column $i$ at time $t$.  On the contrary, in Eq.\eqref{eq:png} the height
$\tilde{h}_i$ relaxes spontaneously even in the absence of deposition to column $i$ at time $t$
because $\tilde{h}_i(t)$ is defined to be the maximum of the triple $\{\tilde{h}_{i-1}(t - 1),
\tilde{h}_i(t - 1), \tilde{h}_{i+1}(t - 1)\}$.  Note that process described by Eq.\eqref{eq:nnn} is
sometimes referred to as ``dynamics with multiplicative noise.''

\subsection{Clusters, crevices, and scaling exponents in~the~growing
  heap}
\label{sect:definitions}

Let us now take a closer look at Fig.~\ref{fig:1}.  We say that two
particles in a heap are \emph{connected} if they touch one another at
corners or if one is situated directly on the top of the other.

It often happens that the upper particle is connected simultaneously
to two lower particles.  For reasons that will become clear shortly,
it is better to avoid these ``one-on-two'' configurations.  The model
is therefore slightly augmented: one assumes in Eqs~\eqref{eq:1b} and
\eqref{eq:nnn} that
\begin{equation}
  \ell = \ell(t) = 1 + 10^{-10}\xi(t),
  \label{eq:modification}
\end{equation}
where $\xi(t)$ are independent normal random variables.  It is clear,
and well supported by numerical experiments, that this modification
removes the possibility of ``one-on-two'' configurations while
preserving, within the limits of statistical errors, statistical
characteristics of the heap for $\ell \equiv 1$.  Alternatively one
might resolve ``one-on-two'' configurations for $\xi=0$ by simply
disconnecting the upper particle from one of its two lower neighbors
at random.  Either way, elimination of one-on-two configurations allow
us to define a unique ``path'' corresponding to every particle, namely
a backward directed chain of connected particles going from a given
particle to the bottom level of the heap.

Consider all connected paths originating from the topmost particles.
These paths can merge.  We define the \textbf{backbone} of a cluster
as the connected set of such paths, i.e., the union of all paths that
end up at the same bottom level particle.  It is easy to see that the
bottom level particles are split into two classes: those that are
reached by the paths originated at the top of the heap and those that
are not.  Obviously the first class gets smaller as $t$ increases.
For every particle from this class define \textbf{cluster} as the
collection of all paths ending up at this particle.  The difference
between a backbone and a cluster is that clusters contain paths not
necessarily originating from the top level particles.

We say that a pair of two top level particles occupying adjacent
columns defines a \textbf{shock}, which is located between them, if
they belong to two different clusters.  The channel of white space
between two neighboring clusters is called a {\bf crevice}.  Clearly
every crevice is associated with a shock at the top, and the connected
paths from top particles defining the shock form the left and right
boundary of a crevice.  Connecting shocks at adjacent time moments, we
get curves that branch forward in time and play a role dual to that of
backbones.  These curves are sketched in Fig.~\ref{fig:1} in black.

It is clear from Fig.~\ref{fig:1} that many channels that are
initially present at bottom of the bulk then merge at some height,
blocking the growth of the clusters situated in between.  Thus
crevices have tree-like structure just as clusters, but contrary to
clusters they merge upward.  This causes the number of percolating
clusters and crevices to decrease as a function of $h$.  In the
thermodynamic limit this behavior is characterized by the following
three scaling exponents whose values are identified in
Section~\ref{sect:scaling}.

The \textbf{thinning exponent} $\alpha$ characterizes the expected
number $\langle c(h) \rangle$ of percolating crevices (or,
equivalently, percolating clusters) at height $h$:
\begin{equation}
  \langle c(h)\rangle \sim h^{-\alpha}
  \label{eq:c}
\end{equation}

The \textbf{wander exponent} $\beta$ characterizes the expected
mean square displacement (in the units of $L$) of the boundary of
percolating cluster between the bottom of the bulk and a specified
height~$h$:
\begin{equation}
  \langle \Delta x^2(h)\rangle \sim h^{\beta}
  \label{eq:r}
\end{equation}

The \textbf{mass exponent} $\tau$ characterizes the mass distribution of clusters:
\begin{equation}
  P(m) \sim m^{-\tau},
\end{equation}
where $P(m)$ is the proportion of clusters of mass~$m$ in the
ensemble.

\section{Ballistic deposition and Burgers turbulence}
\label{sect:3}

\subsection{Variational formulation of the BD}
\label{sect:BDvariational}

The discrete equation
\begin{equation}
  h_i(t) = \max_k\, [h_k(t - 1) - L_{k, i}(t)] + \ell\eta_{i}(t),
  \label{eq:7}
\end{equation}
whose particular cases for specific choices of~$L_{k, i}(t)$ are the
BD model \eqref{eq:nnn} and the discrete PNG model \eqref{eq:png},
admits a natural variational formulation.

Fix some initial condition $h_i(0)$ and consider the discrete
``variational'' problem of finding a trajectory $(\gamma(0),
\gamma(1), \dots, \gamma(t))$ that satisfies the ``boundary
condition'' $\gamma(t) = i$ and maximizes the discrete ``action''
\begin{equation}
  \mathcal A_0^t(\gamma) = h_{\gamma(0)}(0) - \sum_{1\le s\le t}
  [L_{\gamma(s - 1), \gamma(s)}(s) - \ell\eta_{\gamma(s)}(s)].
  \label{eq:10}
\end{equation}
The function in the square brackets plays a role of a discrete
``Lagrangian'' of the system.  The problem bears an obvious
resemblance to the zero temperature limit of the free energy of a
statistical system, expressed as the sum over configurations~$\gamma$:
$$
\lim_{T\to 0} T \ln (e^{\frac{1}{T} A_1} + \dots + e^{\frac{1}{T} A_N}) \to \max \{A_1, \dots,
A_N\}
$$
Another obvious connection is with mechanics, where the dynamical
trajectory can be found by optimizing the corresponding action (in our
case, at variance with the usual convention, the action is
\emph{maximized}).

Action maximization in Eq.~\eqref{eq:10} is related to solving
Eq.~\eqref{eq:7} as follows.  To be specific, consider the BD growth
\eqref{eq:nnn}, where particles are added to the system as ``dropping
events'' $(i(s), s)$ in $(1 + 1)$ dimensional discrete space-time.
Maximization of the action $\mathcal A_0^t$ in~\eqref{eq:10} amounts
to finding a trajectory that terminates at $(i, t)$ and passes through
a maximal number of dropping events under the following constraint:
the trajectory stays constant, $\gamma(s) = \gamma(s - 1)$, unless
$\gamma(s - 1) = i(s) \pm 1$, i.e., there is a dropping event in
adjacent column.  In the latter case the trajectory may (but does
\emph{not} necessarily have to) jump to $i(s)$ at time step~$s$.
Note that for the PNG model \eqref{eq:png} this constraint is relaxed:
a trajectory may jump at all times, but only to adjacent columns.
Otherwise the two models are structurally similar, and the rest of the
argument in this subsection applies to both.

The lack of a strict obligation to pass through an adjacent dropping
event allows to ``collect'' dropping events more efficiently: it is
easy to construct trajectories for which it is more profitable, from
the point of view of maximizing the number of dropping events, to skip
some isolated dropping events in order not to be driven away from a
later series of several adjacent dropping events.

Direct maximization of the action \eqref{eq:10} is a difficult problem
because the solution depends on the whole future history of dropping
events.  Observe however that for all $1\le j\le L$, $1\le s\le t$ the
height function $h_j(s)$ gives the maximal number of dropping events
available for a trajectory coming to the point $(j, s)$, and this fact
can be exploited to construct a maximizing trajectory in
\emph{reverse} time.

Consider again the BD case where $L_{k,j}(s) = L_{k,j}^{\eta_{j}(s)}$.
Then the maximizing trajectory passing through an arbitrary $(i, t)$
can be reconstructed by setting $\gamma(t) = i$ and solving
recursively
\begin{equation}
  \gamma(s - 1) = \arg\max_k\,[h_k(s - 1)
  - L^{\eta_{\gamma(s)}(s)}_{k, \gamma(s)}] + \ell\eta_{\gamma(s)}(s)
  \label{eq:13}
\end{equation}
for $s = t, t - 1, \dots, 1$.  Here $\arg\max_k$ is the standard
notation for the value of $k$ that provides maximum to the expression
in the r.h.s.\ of~\eqref{eq:13}.

The algorithmic implementation of the above goes as follows.  Solve
first Eq.~\eqref{eq:7} ``upstairs'' starting from given initial
conditions and obtain the set of values $h_1(s), h_2(s), \dots,h_L(s)$
for all $0\le s\le t$.  Then choose a specific point, say $(i, t)$,
and restore the path to this point going ``downstairs,'' i.e., back in
time, by solving Eq.~\eqref{eq:13} step by step.  This procedure
defines a trajectory maximizing the action $\mathcal A_0^t$ in
Eq.~\eqref{eq:10}.  This class of algorithms is known in the
optimization theory as \emph{dynamic programming}, and
Eq.~\eqref{eq:7} is called the \emph{Bellman equation} (see, e.g., the
classical book \cite{bellman}).

\begin{figure*}
\epsfig{file=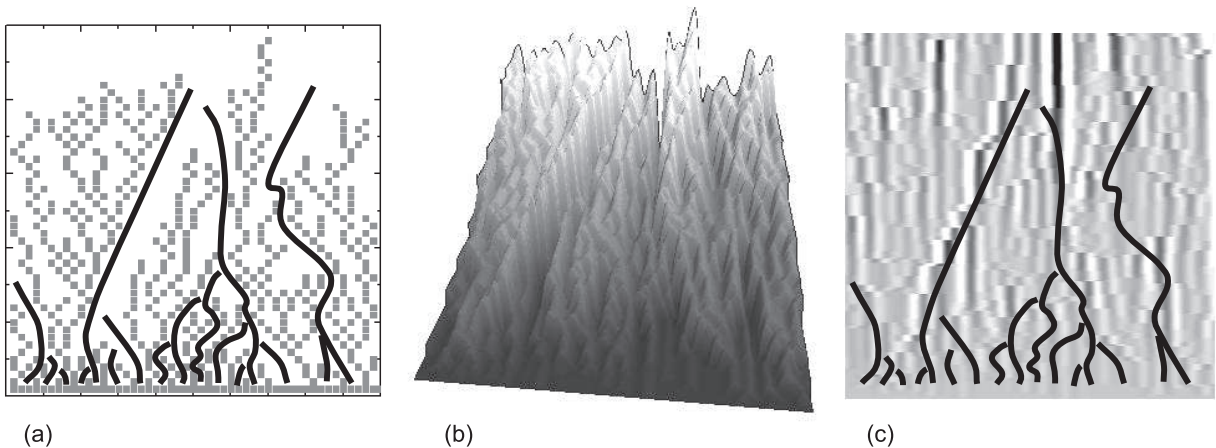,width=16cm}
\caption{(a) Heap growing by sequential deposition with highlighted
  crevices; (b) the growing heap in the $(2 + 1)$ dimensional
  spacetime; (c) density plot of 2nd local difference (discrete analog
  of 2nd derivative) of the height, which highlights the
  discontinuities corresponding to shocks.  Panes (a) and (c)
  represent front and top views, respectively, of the
  three-dimensional structure in pane (b).}
\label{fig:3d}
\end{figure*}

\subsection{BD heaps and the Burgers turbulence}
\label{sect:hje-shocks}

It turns out that there is a far-reaching analogy between the BD
deposition model and phenomenology of ``shocks'' and ``minimizers''
for the Burgers or Hamilton--Jacobi equation with random forcing (see,
e.g., \cite{khanin_bec}).  We first recall the latter.

Consider the inviscid Burgers equation
$$
\partial_t u + u\partial_x u = -\partial_x \eta(x, t),
$$
where $\eta(x, t)$ is the forcing potential.  The substitution $u
= \partial_x h$ transforms this equation into
$$
\partial_t h + (\partial_x h)^2/2 + \eta(x, t) = 0.
$$
More generally, one can consider the Hamilton-Jacobi equation
\begin{equation}
  \partial_t h + H(\partial_x h) + \eta(x, t) = 0,
  \label{eq:hje}
\end{equation}
where $H(p)$ is a convex function representing the kinetic energy.
Using the Legendre transform representation $H(p) = \max_v [pv -
L(v)]$, one can write
$$
\partial_t h + v\,\partial_x h - L(v) + \eta(x, t) \le 0
$$
with equality only for $\partial_x h = L'(v)$, i.e., $v =
H'(\partial_x h)$.  Hence along any trajectory $\gamma(t)$ the rate of
change of~$h$ is bounded by the Lagrangian
$$
\frac d{dt} h(\gamma, t) \le L(\dot\gamma) - \eta(\gamma, t)
$$
(here $\dot\gamma = d\gamma/dt$), which implies for any $\gamma$
passing through $x$ at time $t$ that
\begin{equation}
  h(x, t) \le \mathcal A_0^t[\gamma]
  = h(\gamma(0), 0) + \int_0^t [L(\dot\gamma) - \eta(\gamma, s)]\, ds
  \label{eq:10c}
\end{equation}
with equality only for \textbf{minimizers} of the action, which must
satisfy the equation
\begin{equation}
  \dot\gamma(t) \equiv H'(\partial_x h(\gamma, t)).
  \label{eq:13c}
\end{equation}

The Hamilton--Jacobi equation \eqref{eq:hje} is thus intimately
connected with the variational problem of minimizing the action
\eqref{eq:10c}, just as the Bellman equation \eqref{eq:7} arises in
maximization of the discrete ``action'' \eqref{eq:10}.  Note in
particular the similar structure of the action (the difference in sign
results in maximization replacing minimization in the discrete case).
Moreover, a known solution $h$ to \eqref{eq:hje} allows to reconstruct
minimizing trajectories using \eqref{eq:13c}, much as \eqref{eq:13}
generates maximizing trajectories in the discrete problem.

It is therefore natural to consider the discrete maximizing trajectories defined in the previous
subsection as analogs of continuous minimizers.  There is one apparent difference: continuous
minimizers never cross, while discrete maximizing paths merge and form tree-like structures.
However continuous minimizers have a tendency to approach each other with exponential rate in
reverse time due to hyperbolicity, and in the discrete case the same hyperbolicity manifests itself
in the exponentially decreasing probability for two adjacent maximizers to stay separate as time
runs backwards.

We are now in position to establish the relation between discrete
maximizers and connected paths defined within the heap in
Section~\ref{sect:2}.  Lift the maximizing trajectories to the $(i, t,
h)$ space by setting $h = \mathcal A_0^t(\gamma)$ for a maximizer
$\gamma$ such that $\gamma(t) = i$.  Then connected paths are given by
the projection of these ``lifted'' maximizers to the $(x, h)$ plane
(see Fig.~\ref{fig:3d}).  In other words, the intervals of time
between successive dropping events along a maximizer are collapsed
into unit steps in~$h$.  Correspondingly the transversal fluctuations
of maximizers as a function of time are transformed to transversal
fluctuations of connected path as a function of height $h$.

The analogy between continuous minimizers and discrete maximizers
extends to shocks.  In the Burgers turbulence it typically happens
that two or more minimizing trajectories, which start at different
initial locations, pass through same point~$x$ at time~$t$, so that
the map from $(x, t)$ to the initial location is discontinuous (see,
e.g., \cite{khanin_bec}).  These discontinuities are called
\textbf{shocks}; in spacetime they form continuous shock curves.  This
definition is obviously parallel to the definition of shocks given in
the BD setting in Section~\ref{sect:2} (and has inspired the latter).

\section{From BD patterns to Airy processes}
\label{sect:airy}

\subsection{Basics of classical KPZ scaling}
\label{sect:kpz}

Recall first the basics of classical KPZ scaling related to $Airy_1$
process, which is closest to our setting.  The scheme described below
is due to Sasamoto~\cite{sasam}.

Consider a directed random walk on a $(1 + 1)$ dimensional lattice.
Suppose that the space-time lattice is equipped by a random potential
with independent values $\eta_i(s)$ at each point~$(i, s)$.  Then for
every $i$ one can consider the maximum of an action over all random walk paths of
length $t$ terminating at that point, i.e., define
$$
a_i(t) = \max_{\gamma\colon \gamma(t) = i} \sum_{1\le s\le t} \eta_{\gamma(s)}(s).
$$
where $L_{k, i}$ is a ``kinetic'' part of the action that ensures a
certain control of how far the trajectory~$\gamma$ can jump over unit
time steps.  It is easy to see that $t^{-1}\, a_i(t) \to c$ at $t \to
\infty$, where $c$ is some nonrandom constant independent of $i$.  We
now consider the rescaled process
\begin{equation}
A_t(x) = \frac 1{\beta t^{1/3}}\, (a_{\alpha t^{2/3}x}(t) - ct).
\label{eq:airy1}
\end{equation}

The main statement is that $A_t(x)$ converges as $t \to \infty$ to a
universal spatially homogeneous limit process called $Airy_1(x)$.
Universality here means that whenever one optimizes in a disordered
medium the action of a path from a point that varies over a line to a
parallel line separated from the first one by distance $t$
(``point-to-line last-passage percolation''), the process
corresponding to the optimal action converges as $t \to \infty$ to the
$Airy_1$ process.

Note that spatial homogeneity of $Airy_1(x)$ immediately follows from
the construction.  Of course one has to ensure convergence by
subtracting the mean value of order $t$, normalizing the difference by
$t^{1/3}$ and rescaling the starting point by~$t^{2/3}$.  The
constants $\alpha$ and $\beta$ in~\eqref{eq:airy1} are nonuniversal
and should be chosen properly to ensure convergence to the standard
Airy process.  A~similarly rescaled ``point-to-point'' percolation
results in the $Airy_2$ process.

\subsection{The Airy process for BD pattern}
\label{sec:airy-process-bd}

As we have shown in Section~\ref{sect:3} the height function in the BD
process can be viewed as given by maximization procedure for random
paths in random potential.  The only difference with the classical
picture just described is related to the rarity of the deposition
events.  In other words, in order to achieve the displacement of order
1 in space direction one needs time of order~$L$.  This explains why
time has to be rescaled.

The most natural way to do this is through a local stochastic change
of time variable.  Namely we collapse the time between two deposition
events to~$1$. This is exactly the transformation from (lifted)
maximizers to connected paths presented in Section~\ref{sect:3}.  It
is therefore no surprise that the $Airy_1$ process can be obtained
from the BD height function:
\begin{equation}
   \lim_{t\to\infty} \frac 1{\beta (t/L)^{1/3}}
    (h_{\alpha (t/L)^{2/3}x}(t) - t/L) = Airy_1(x).
 \label{eq:airy1_2}
\end{equation}
This formula simply indicates that the appropriately rescaled height
function in BD is the visualization of the process which converges in
the thermodynamic limit to the Airy process.

\subsection{The ``hairy Airy'' process}
\label{sec:hairy-airy-process}

The $Airy_1$ process carries only part of the information about the
system: it is oblivious to the maximizing trajectory associated to the
(rescaled) point~$(x, t)$.  It is therefore natural to consider the
limit
\begin{multline}
  \label{eq:airy2}
  \left(\frac {a_{\alpha t^{2/3}x}(t) - ct}{\beta t^{1/3}},\
    \frac {\gamma_{\alpha t^{2/3}x,
        t}(ts)}{\alpha t^{2/3}}\right) \\
  \xrightarrow[t\to\infty]{} (Airy_1(x), \Gamma_x(s)),
\end{multline}
where $\gamma_{i, t}$ is the maximizing trajectory that passes
through~$i$ at time~$t$ and $\Gamma_x(s)$ is a continuous path defined
over $[0, 1]$ such that $\Gamma_x(1) = x$.  We call this limit the
\emph{equipped Airy process}.

As just before, in the BD setting we collapse time intervals between
adjacent deposition events to unit steps and get particle paths
instead of maximizing trajectories in formula~\eqref{eq:airy2} above.
Applying transversal rescaling $\alpha (t/L)^{2/3}x$ and height
rescaling $\beta(t/L)^{1/3}$ as in~\eqref{eq:airy1_2}, we get a
realization of equipped Airy process from the rescaled BD heap.  This
process describes the joint distribution of fluctuations of the height
function and transversal displacements of cluster boundaries in the
spatially homogeneous BD process.

\begin{figure}[ht]
\epsfig{file=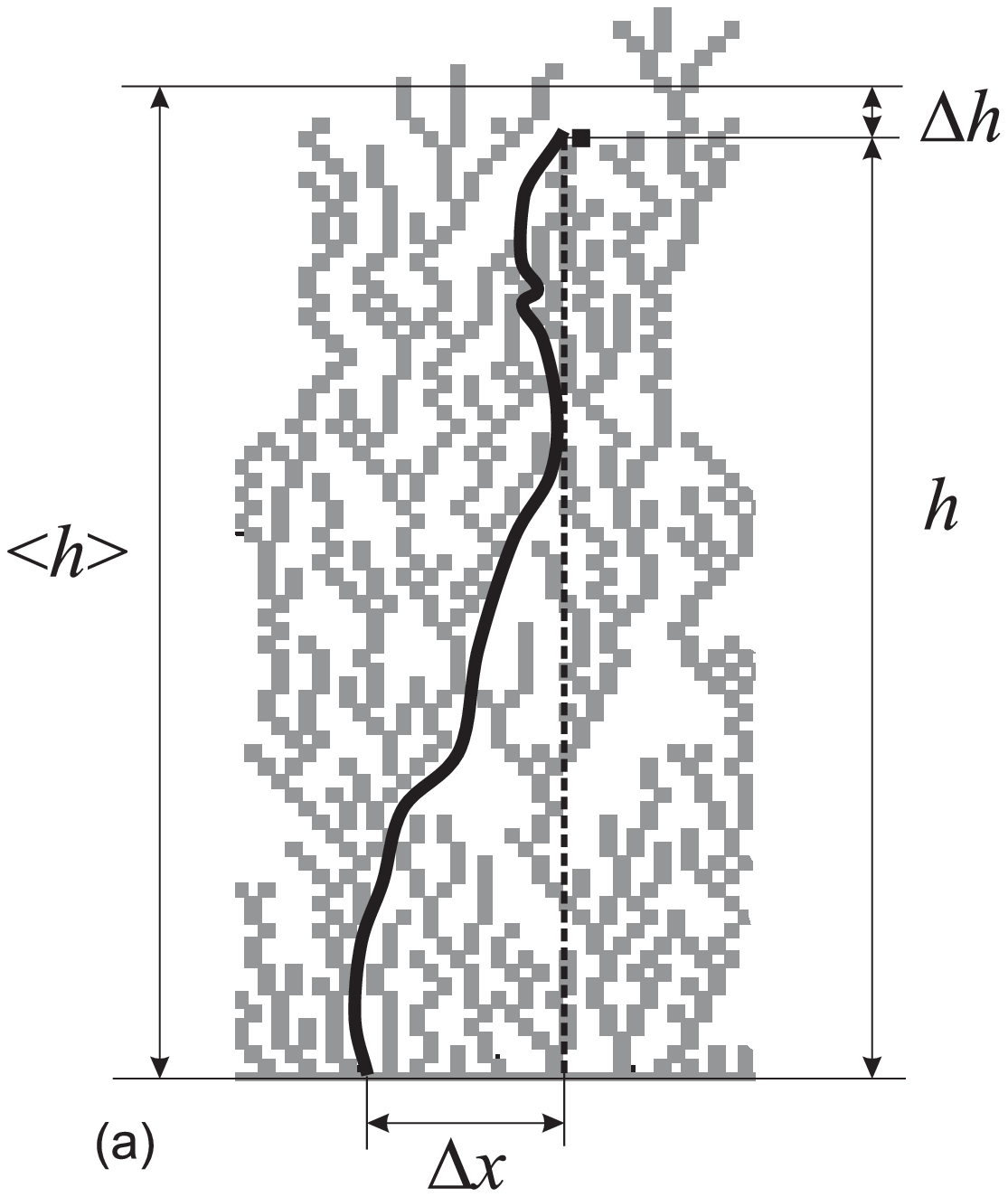,width=6cm} \\
\hspace{-0.3cm} \epsfig{file=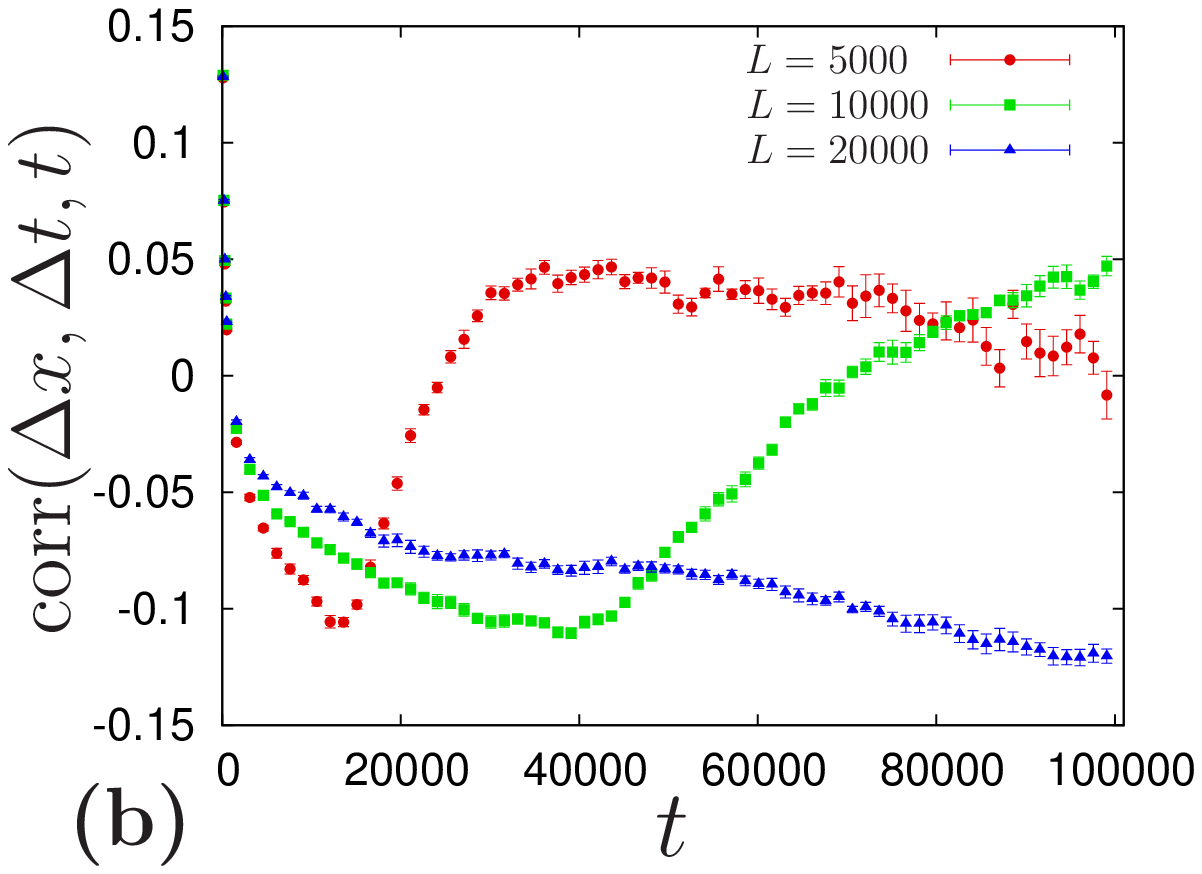,width=6.15cm} \\
\epsfig{file=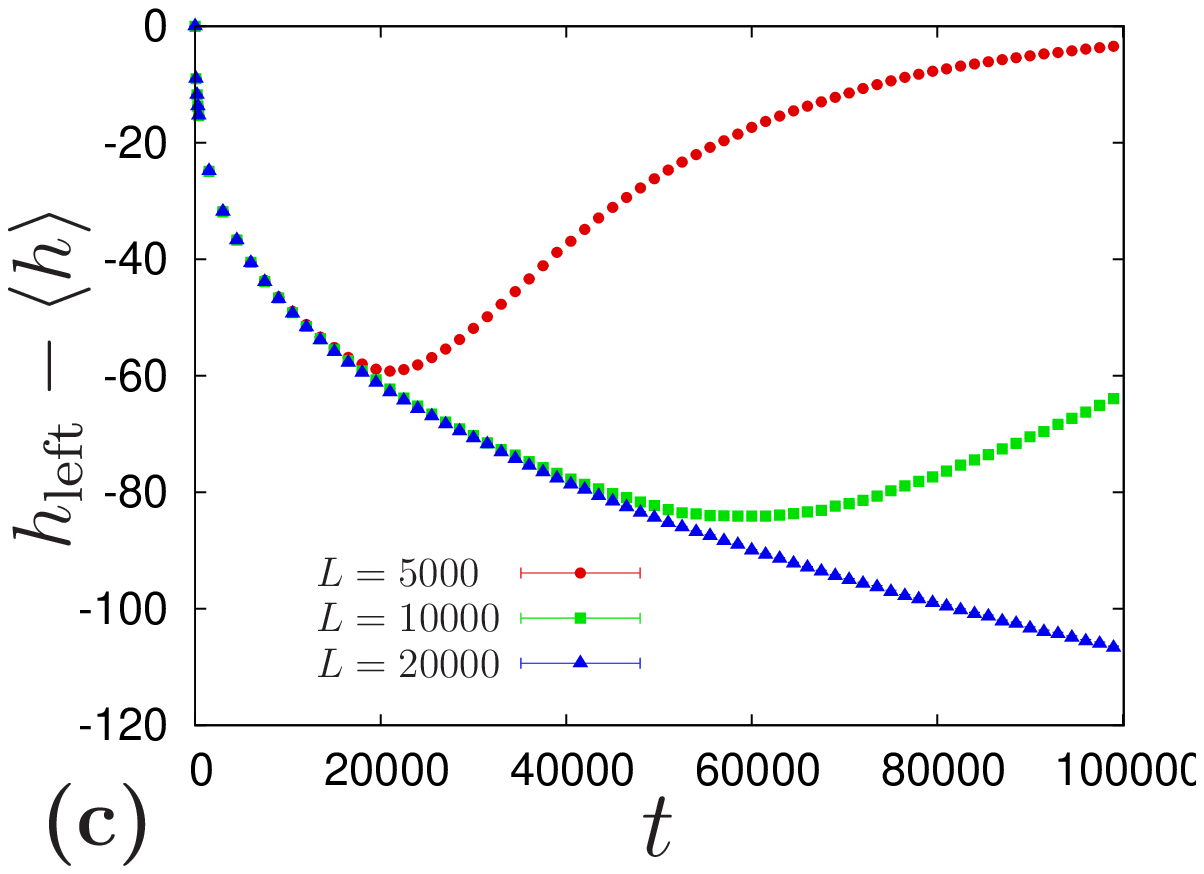,width=6cm}
\caption{(Color online) a) Correlation between the height of cluster's
  boundary and the displacement of the corresponding connected path;
  b) the corresponding correlation coefficient; c) the averaged
  difference between the height of cluster's right boundary and the
  mean height of the BD growing interface.}
\label{fig:what_corr}
\end{figure}

In other words, the rescaled height function for the BD model alone is a realization of the
$Airy_1$ process, while the rescaled height function together with the rescaled forest of
maximizers corresponds to a realization of the equipped Airy process.  The distinctive geometric
features of this joint process suggests the name ``\textbf{hairy Airy process},'' cf.\
Fig.~\ref{fig:what_corr}a.

We demonstrate the existence of correlations in the joint distribution
for the hairy Airy process by computing numerically the joint
distribution of the height fluctuation $\Delta h$ at the top of a
shock and the corresponding displacement $\Delta x$ of the cluster
boundary.  To be precise, we compute the correlation coefficient
between the fluctuations of the displacement $\Delta x$ of the right
boundary of a cluster (or equivalently a backbone) and the height
fluctuation $\Delta h$ at the top right point of the same cluster, see
Fig.~\ref{fig:what_corr}a.  For convenience we explicitly recall here
the standard definition of the correlation coefficient
$\mathrm{corr}\{a,b\}$ between two random variables $a$ and $b$:
\begin{equation}
  \mathrm{corr}\{a, b\}
  = \frac{\langle(a - \langle a\rangle)\, (b - \langle b \rangle)}
  {\sqrt{\langle(a - \langle a \rangle)^2\rangle\,
      \langle(b - \langle b \rangle)^2\rangle}}.
  \label{eq:corr}
\end{equation}

Let the top right particle of some cluster be located at time~$t$ in
column~$j$.  Let $h_j(t)$ be its height and $\langle h(t) \rangle$ the
mean height of the whole surface at time $t$.  Denote furthermore
$h_j(t) - \langle h(t)\rangle$ by $\Delta h_j(t)$ and the displacement
of the right boundary of the same cluster at time $t$, measured from
the position of this boundary at $t=0$, by~$\Delta x_j(t)$.  We fix a
time~$t$, collect for each cluster the joint information $(\Delta
h_j(t), \Delta x_j(t))$, and perform averaging over all clusters.
Behavior of the corresponding time-dependent correlation coefficient
$\mathrm{corr}\{\Delta h, \Delta x\}$, where the angle brackets
correspond to averaging over the sample, is shown in
Fig.~\ref{fig:what_corr}b for different time values.

Strong correlations between the vertical and horizontal displacements
of cluster boundaries are clearly seen in the data.  The negative sign
of these correlations is due to the fact that the height of the top
right particle in a typical cluster is smaller than the averaged
height of the growing BD interface.  This observation is supported by
Fig.~\ref{fig:what_corr}c, where the averaged difference between the
height of the cluster right boundary and the averaged height of the
interface is plotted against time.  Clearly this difference is always
negative and tends to~$0$ from below as $t\to \infty$.  One may
speculate that growth of the left- and ritghmost connected paths in a
cluster is slower due to screening between neighboring clusters.

\begin{figure}[ht]
\epsfig{file=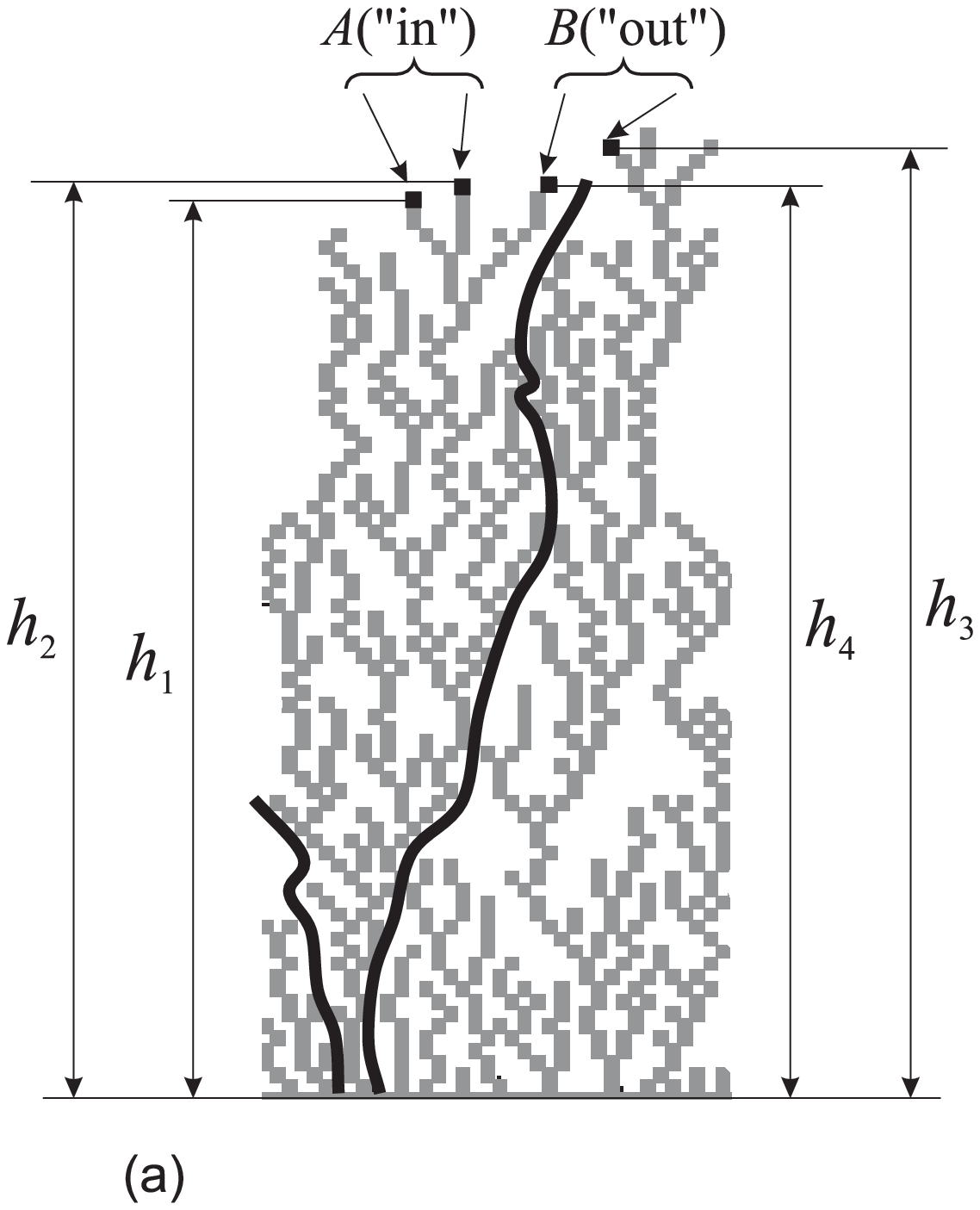,width=6cm} \\ \epsfig{file=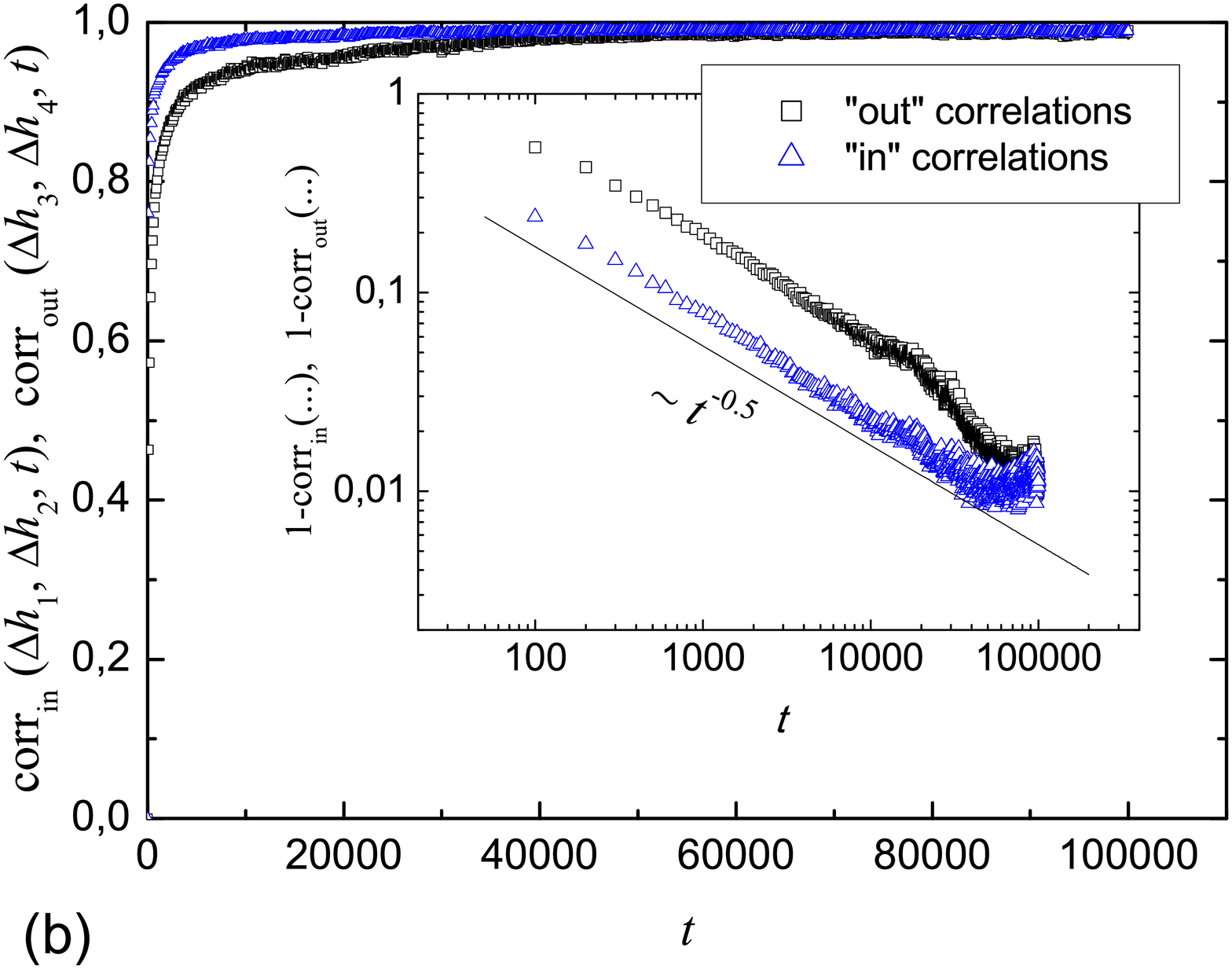,width=7cm}
\caption{(Color online) a) Correlation between the heights inside the
  cluster (``in'') and separated by a shock (``out''); b) the
  corresponding correlation coefficients.}
\label{fig:correlator}
\end{figure}

In order to better understand the influence of clusters on the
morphological structure of the growing BD surface, we also compute the
joint distribution of height fluctuations in two columns separated by
distance $\delta = 3$ in lattice units, as shown in
Fig.~\ref{fig:correlator}a.  Two different situations are
distinguished: i) two test column belong to the same cluster
(configuration $A$), and ii) two test columns belong to different
clusters, i.e., are separated by a shock (configuration $B$).

Computing the correlation coefficient $\mathrm{corr}\{\Delta h_k,
\Delta h_m\}$ according to~\eqref{eq:corr}, we see that correlations
between $\Delta h_1$ and $\Delta h_2$ inside a cluster are stronger
than those across a shock between different clusters.

\section{Scaling analysis of BD patterns}
\label{sect:scaling}

\subsection{``Thinning'' of clusters and wandering of their boundaries}
\label{sec:clust-dens-crev}

Relying on the connection between shocks and boundaries of clusters,
we can directly transfer the scaling arguments of statistics of shocks
developed in~\cite{khanin_bec} to the scaling analysis of a growing BD
heap and determine the values of the scaling exponents $\alpha$ and
$\beta$ in the dependencies $c(h)\sim h^{-\alpha}$ and $\Delta
x^2(h)\sim h^{\beta}$ defined correspondingly in Eqs~\eqref{eq:c}
and~\eqref{eq:r}.  Recall that $c(h)$ is the averaged number of
clusters percolating to height~$h$ and $\Delta x^2(h)$ is the mean
square displacement of a cluster boundary at height~$h$.

Denote by $d(t)$ the horizontal size of a cluster at time~$t$.  At $t
= 0$ the cluster has zero size, i.e., $d(0) = 0$.  In what follows we
shall use the obvious fact that the growth time~$t$ in the sequential
deposition process is proportional to the average height~$h$ of the
growing heap and, consequently, to the cluster height --- see, for
example, Fig.~\ref{fig:3d}b.

The typical value of $d(h)$ can be obtained by scaling considerations.
Namely, growth of $d(h)$ is determined by two additive effects.  On
the one hand, there is a ``driving force'' promoting the ``smearing''
of the cluster due to the velocity fluctuations.  For BD this effect
can be estimated as follows.  Consider clusters with size of order
$d$.  Under the uniform random ``rainfall'' of deposited particles,
one cluster can randomly screen part of its neighbors, and increase
its own ``spot.''  Since different clusters are correlated weakly, it
is natural to conjecture that the typical scale of fluctuations of
cluster sizes is of order of $\sqrt d$.  Thus the rate $v$ of cluster
``smearing'' due to these fluctuations is $v\sim d/\sqrt{d} \sim
d^{-1/2}$.  Speaking more carefully, the above means that the average
growth rate of the cluster of size $d$ is
$$
v=\frac 1d \sum_{k\le j\le k + d} (h_{j + 1}-h_j),
$$
where $k$ and $k + d$ are the left and the right boundaries of some
cluster.  The increments of $h_j$ are uncorrelated for the uniform
ballistic ``rain'' and $\langle h_{j+1}-h_{j} \rangle = 0$.  It is
therefore natural to expect that $v\sim d^{-1/2}$ as conjectured.

On the other hand, there is ``smearing'' of clusters due to the random
deposition of new particles near the cluster boundary.  This process
can be interpreted as ``diffusion'' of the boundary.  Over time $t$
this diffusion leads to the smearing of the cluster's horizontal size
on typical scale of order of $\sqrt{t}$.

The typical size of a growing cluster at time~$t$ is determined by
additive contributions of these two effects:
\begin{equation}
d(t)\simeq t v + t^{1/2} = \frac{t}{\sqrt{d(t)}} + t^{1/2}
\label{eq:opt}
\end{equation}
The dominant contribution to $d(t)$ comes from the first term, which
is consistent with the physical intuition.  Hence,
\begin{equation}
  d(t) \sim t^{2/3}
  \label{eq:opt2}
\end{equation}
Since $t\sim h$, we immediately come to the conclusion that $d(h) \sim
h^{2/3}$.  (This estimate is a direct paraphrase of the arguments
provided in \cite{khanin_bec} for scaling analysis of statistics of
shocks in the $(1 + 1)$ dimensional Burgers equation with random
forcing).

The density $c(h)$ of independent clusters surviving up to the height
$h$ is inversely proportional to the cluster size, $c(h)\sim
[d(h)]^{-1}$.  Thus,
\begin{equation}
  c(h) \sim h^{-2/3},
  \label{eq:opt3}
\end{equation}
which gives $\alpha = 2/3$.

Furthermore, the typical horizontal mean square displacement $\langle
\Delta x^2(h)\rangle$ of a cluster boundary at height~$h$ can be
estimated simply as
\begin{equation}
  \langle \Delta x^2(h)\rangle = d^2(h) \sim h^{4/3},
  \label{eq:opt4}
\end{equation}
which gives $\beta=4/3$.

\subsection{Mass distribution of clusters}
\label{sec:mass-distr-clust}

\begin{figure}[ht]
\epsfig{file=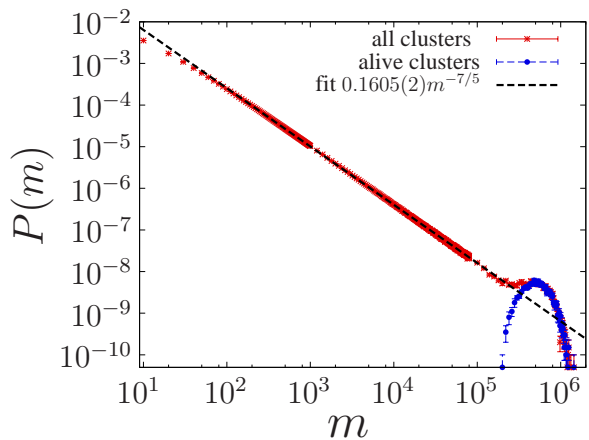,width=8cm}
\caption{The probability $P(m)$ to find in a large system a cluster of mass $m$.}
\label{fig:6}
\end{figure}

This Section contains the scaling analysis of the probability $P(m)\sim m^{-\tau}$ to find a
cluster of mass $m$ in a large aggregate.  To begin with, note that the number of particles, i.e.,
the ``mass'' $m(h)$ of a cluster percolating to height~$h$ can be obtained integrating the
horizontal size, $d(h)$, of cluster at a given height:
\begin{equation}
  m(h) %= \int_0^h d(h')\, \mathrm{d}h'
  \sim \int_0^h (h')^{\alpha}\, \mathrm{d}h' \sim h^{\alpha+1}.
  \label{eq:m}
\end{equation}
From \eqref{eq:opt3} we know that the cumulative probability of clusters surviving until height~$h$
is of the order of $h^{-\alpha}$. This implies that the probability density of clusters at
height~$h$ scales as $h^{-(\alpha+1)}\, \mathrm d h$.  In order to calculate mass distribution, we
change variables from~$h$ to~$m$, take into account that $\mathrm d m \sim h^{\alpha}\, \mathrm d
h$, or $\mathrm d h \sim h^{-\alpha}\, \mathrm d m = m^{-\alpha/(\alpha+1)}\, \mathrm d m$, and get
\begin{equation}
\begin{array}{rcl}
  \label{eq:g2}
  P(m) & \sim & m^{-\tau} \, \mathrm d m
  \sim h^{-(\alpha+1)}\, \mathrm d h
  \sim m^{-1} \, m^{-\alpha/(\alpha+1)}\, \mathrm d m \medskip \\
  & = & m^{-(2\alpha+1)/(\alpha+1)}\, \mathrm d m.
\end{array}
\end{equation}
For $\alpha=2/3$ we get $\tau=7/5$. The exponent $\tau=7/5$ is well supported by our own numerical
simulations shown in Fig.~\ref{fig:6}, and, as mentioned in Introduction, it has been found in
independent laboratory experiments on cluster formation in quasi-two-dimensional electrochemically
formed silver branching structures \cite{silver}.

One can say that he clusters are ranked (ordered) according to their masses and $m$ is the
corresponding rank. Thus Eq.\ \eqref{eq:g2} has similarity with the Zipf's law that appears in many
areas of science ranging from word statistics in linguistics \cite{zipf} to nuclear
multifragmentation \cite{ma,campi} where clusters have power-law distribution in sizes (masses,
charges etc.).

\section{Conclusion}
\label{sect:conclusion}

In this paper we analyze the internal structure of the heap formed in
the course of standard homogeneous ballistic deposition with
next--nearest--neighboring (NNN) interactions in a box.  We have paid
the most attention to the statistics of clusters and the channels
(crevices) separating them.  We have demonstrated that the BD process
can be naturally described in terms of ``dynamic programming''
language associated with the so-called Bellman equation.  The
``dynamic programming'' point of view allows systematic translation of
the study of clusters and crevices in the NNN ballistic deposition
into the language of maximizers and shocks in discrete equations of
the Burgers or Hamilton--Jacobi type.  This is the key point of our
work.  A detailed examination of the corresponding continuous limit
will be the subject of a forthcoming publication.

In particular, the results of the work \cite{khanin_bec} concerning
the statistics of shocks in $(1+1)$ dimensional Burgers turbulence
with random forcing allow the direct interpretation for statistics of
cluster's boundaries (crevices) of growing heap.  This connection
between shocks and crevices has permitted us to compute the scaling
exponents $\alpha$ ($\alpha=2/3$) in the dependence $\langle c(h)
\rangle \sim h^{-\alpha}$, where $\langle c(h) \rangle$ is the average
number of clusters, surviving up to height $h$ and $\beta$
($\beta=4/3$) for the mean square displacement $\langle \Delta
x^2(h)\rangle \sim h^{\beta}$ of crevices as a function of the height
$h$ of the heap.

We have also extended the scaling analysis to the computation of the critical exponent $\tau$
($\tau=7/5$) of the mass distribution of clusters, $P(m) \sim m^{-\tau}$, where $P(m)$ is the
probability density of clusters of mass~$m$ (see Fig.~\ref{fig:6} for comparison of numerical
simulation with scaling dependence \eqref{eq:g2}).  The exponent $\tau$ coincides with the one
found in real experiments on cluster formation in quasi--two dimensional electrochemically formed
silver branching structures \cite{silver}.

The investigation of the morphological structure of surface of the
growing heap splitted in clusters, has lead us to the definition of a
new ``equipped'' Airy process for BD, named the ``hairy Airy
process.''  In our preliminary investigation we have analyzed
numerically its two-point correlation function and have shown the
existence of essential correlations between the fluctuations of the
displacement $\Delta x$ of the cluster's left boundary and the
height's fluctuation $\Delta h$ in the top point of the same cluster's
left boundary location, see Fig.~\ref{fig:what_corr}a.

We believe that the described connection between crevices and shocks
could be a useful tool for deeper understanding of both topics, NNN
ballistic deposition and Burgers turbulence.  For NNN ballistic growth
we could apply the machinery developed in turbulence, while for
turbulence we could use NNN ballistic deposition for straightforward
visualization of some complex chaotic behavior.

Let us end up by noting that many important and puzzling questions
concerning the growth of the heap have not been touched in this paper.
For instance, we have not discussed the question mentioned in the
Introduction: why the bulk density of the heap does not coincide with
the density of local maxima of the growing surface.  Our guess is that
the discrepancy between these densities, $\rho_{\rm bulk}$ and
$\rho_{\rm surf}$ is due to the presence of crevices in the heap.
Another example remaining almost without the attention deals with the
consideration of aging in the growing heap.  The investigation of the
correlation between two heights inside a cluster and separated by the
crevice considered in Fig.~\ref{fig:correlator} gives some hint about
the aging of the heap, however we have not considered the correlation
between two heights separated by the crevice of finite depth.

\begin{acknowledgments}
  We are grateful to G. Carlier for drawing our attention to the fact
  that both the NNN and PNG evolution processes can be expressed in
  terms of a Bellman equation. K.~Khanin and A.~Sobolevski are
  partially supported by the joint CNRS--RFBR project 07--01--92217;
  the latter author also acknowledges the support of the French Agence
  Nationale de la Recherche via project BLAN 07--01--0235 OTARIE.
\end{acknowledgments}

\end{document}